\documentclass[twocolumn]{aastex6}

\usepackage{ulem}
\usepackage{natbib}
\usepackage{amsmath}
\usepackage{color}

\newcommand{\tone}{t_{\rm 1D}}
\newcommand{\tthr}{t_{\rm 3D}}

\begin{document}

\title{High-resolution hydrodynamic simulation of tidal detonation of
  a helium white dwarf by an intermediate mass black hole}

\author{Ataru Tanikawa\altaffilmark{1,2}}

\altaffiltext{1}{Department of Earth Science and Astronomy, College of
  Arts and Sciences, The University of Tokyo, 3-8-1 Komaba, Meguro-ku,
  Tokyo 153-8902, Japan; tanikawa@ea.c.u-tokyo.ac.jp}
\altaffiltext{2}{RIKEN Advanced Institute for Computational Science,
  7-1-26 Minatojima-minami-machi, Chuo-ku, Kobe, Hyogo 650-0047,
  Japan}

\begin{abstract}

We demonstrate tidal detonation during a tidal disruption event (TDE)
of a helium (He) white dwarf (WD) with $0.45M_\odot$ by an
intermediate mass black hole (IMBH) by extremely high-resolution
simulations. \cite{2017ApJ...839...81T} have showed tidal detonation
in previous studies results from unphysical heating due to
low-resolution simulations, and such unphysical heating occurs in
3-dimensional (3D) smoothed particle hydrodynamics (SPH) simulations
even with $10$ million SPH particles.  In order to avoid such
unphysical heating, we perform 3D SPH simulations up to $300$ million
SPH particles, and 1D mesh simulations using flow structure in the 3D
SPH simulations for 1D initial conditions. The 1D mesh simulations
have higher resolution than the 3D SPH simulations. We show tidal
detonation occurs, and confirm this result is perfectly converged with
different space resolution in both 3D SPH and 1D mesh simulations. We
find detonation waves independently arises in leading parts of the WD,
and yield large amounts of $^{56}$Ni. Although detonation waves are
not generated in trailing parts of the WD, the trailing parts would
receive detonation waves generated in the leading parts, and would
leave large amounts of Si group elements. Eventually, this He~WD~TDE
would synthesize $^{56}$Ni of $0.30M_\odot$ and Si group elements of
$0.08M_\odot$, and could be observed as a luminous thermonuclear
transient comparable to type Ia supernovae.

\end{abstract}


\section{Introduction}
\label{sec:introduction}

A tidal disruption event (TDE) is a phenomenon in which a star is
tidally disrupted by a black hole (BH). TDEs have luminosities powered
by accretion of stellar debris onto a BH. So far, many TDEs have been
found \citep[see reviews
  by][]{2015JHEAp...7..148K,2017ApJ...838..149A,2018arXiv180110180S}. In
most of these TDEs, main sequence (MS) stars are disrupted by massive
black holes (MBH) with $10^6M_\odot$ to $10^8M_\odot$. These TDEs can
be useful for measuring physical quantities of MBHs, such as mass and
spin.

Studies of MS~TDEs are aimed at MBHs, whereas TDEs of white dwarfs
(WDs) can contribute to studies of intermediate mass black holes
(IMBHs). Similarly to MS~TDEs by IMBHs, WD~TDEs by IMBHs have bright
flares driven by accretion of their debris onto the IMBHs
\citep{2010MNRAS.409L..25Z,2011ApJ...726...34C,2012ApJ...749..117H,2014PhRvD..90f4020C,2014ApJ...794....9M,2015ApJ...804...85S,2016ApJ...833..110I,2017ApJ...841..132L}.
Moreover, luminosities of WD~TDEs could be powered by radioactive
decay of nuclei synthesized by tidal detonation
\citep[e.g.][]{1989A&A...209..103L}. The tidal detonation could occur
as follows. During a TDE, a WD is elongated in the direction of the
orbital plane (hereafter, $x$-$y$ plane), and however is compressed in
the direction perpendicular to the $x$-$y$ plane (hereafter,
$z$-direction). The compression heats the WD, and triggers explosive
nuclear reactions. Finally, the explosive nuclear reactions synthesize
large amounts of radioactive nuclei
\citep{2008CoPhC.179..184R,2009ApJ...695..404R}.
\cite{2016ApJ...819....3M} have showed the luminosity of the
synthesized radioactive nuclei at peak is larger than the
accretion-powered luminosity (the Eddington luminosity of an IMBH) by
two orders of magnitude.  WD~TDEs can launch jets whose luminosities
are much larger than the radioactive luminosity if observers are along
with the jet axis
\citep{2011MNRAS.417L..51V,2013A&A...552A...5V,2011ApJ...743..134K,2012ApJ...749...92K,2014MNRAS.437.2744T,2014ApJ...794....9M},
although the jet luminosities strongly depends on line-of-sight
directions.  WD~TDEs also emit gravitational wave (GW) radiation,
although the GW frequency is fit to space-based GW detectors
\citep{2014ApJ...795..135E}.  Current and future optical surveys can
find many IMBHs, searching for tidal detonation of WDs. The abundance
of IMBHs will be an important key to reveal the formation process of
MBHs.

\cite{2017ApJ...839...81T}, hereafter Paper~I, have revisited tidal
detonation of a WD. Paper~I has shown adiabatic compression cannot
heat a WD sufficiently for tidal detonation in the following reason.
A WD should be adiabatically compressed by at least $4$ orders of
magnitude in order to experience tidal detonation. However, such
adiabatic compression is impossible. A WD can approach an IMBH without
swallowing nor swallowed by the IMBH when a penetration factor $\beta$
is less than $20$
\citep{1989A&A...209..103L,2009ApJ...695..404R,2017arXiv170505526K},
where $\beta$ is defined as the ratio of a tidal disruption radius to
a pericenter distance. The size of a WD in the $z$-direction, $z_{\rm
  min}$, can be written as $z_{\rm min}/R_{\rm wd} \sim \beta^{-3}$ at
the pericenter \citep{2013MNRAS.435.1809S}, where $R_{\rm wd}$ is the
original radius of the WD. Hence, a WD is compressed by at most a
factor of $8000$, and in reality is less compressed, since it is
elongated in the direction of $x$-$y$ plane.
\cite{1989A&A...209..103L} have argued a helium (He) WD can experience
tidal detonation by adiabatic compression. However, their He~WD has
unrealistically high density, $\sim 10^7$~g~cm$^{-3}$, or
unrealistically high mass, $0.6M_\odot$. Note that the upper limit of
He~WD mass is about $0.5M_\odot$ \citep{2017MNRAS.470.4473P}. If a
He~WD has such high density, tidal detonation can occur when the He~WD
is compressed only by two orders of magnitude.
\cite{2004ApJ...610..368W} have suggested compression of a WD causes
the central density to exceed the threshold for pycnonuclear
reactions. However, they have overestimated the compression, since
they have not taken into account stretch of the WD by a tidal field.

Thus, tidal detonation requires shock compression. A shock wave can
arise when $\beta$ is sufficiently larger than unity
\citep{2004ApJ...615..855K}. In the case of $\beta < 12$, where most
of WD~TDEs occur, the mechanism of shock generation is as follows
\citep{2008A&A...481..259B}.  A WD approaches to an IMBH, and is
compressed in the $z$-direction. At some time, the central pressure of
the WD increases instantaneously. Then, the WD bounces back. The
bounce generates a pressure wave propagating outward along the
$z$-direction. The pressure wave steepens into a shock wave near the
WD surface.

However, the generation of a shock wave is not sufficient condition
for tidal detonation. The shock wave has to raise temperature so
highly that nuclear reactions start. Moreover, the nuclear reactions
have to be explosive. Otherwise, they cease soon.

In this paper, we investigate whether a shock wave arising during a
WD~TDE triggers tidal
detonation. \cite{2008CoPhC.179..184R,2009ApJ...695..404R} have
reported tidal detonation occurs in He~WDs and CO~WDs by 3-dimensional
(3D) smoothed particle hydrodynamics (SPH) simulations. However, they
have not shown shock generation in their simulations
explicitly. Paper~I have shown tidal detonation in Rosswogs'
simulations is triggered by spurious heating due to failure to resolve
the height of a WD from the $x$-$y$ plane. Paper~I have demonstrated
high-resolution 1-dimensional (1D) mesh simulations in which shock
compression generates tidal detonation. Here, initial conditions of
the 1D mesh simulations are WD structure along with the $z$-direction
which is extracted from our 3D SPH simulations. However, in Paper~I,
our method to make 1D initial conditions is not sophisticated. Density
profiles in the 1D mesh simulations tend to be higher than in the 3D
SPH simulations during the evolution. This is because 3D effects, such
as a tidal field, are ignored in the 1D mesh simulations. Since
nuclear reactions become active under higher density environment,
shock compression is easier to trigger tidal detonation in the 1D mesh
simulations than in reality. Therefore, Paper~I have not confirmed
whether tidal detonation occurs or not. Moreover, the overestimate of
density affects nucleosynthesis if tidal detonation occurs.

For this purpose, we develop a method to make 1D initial conditions
extracting WD structure along with the $z$-direction from 3D SPH
simulations. Owing to this method, density evolution in 1D mesh
simulations is the same as in 3D SPH simulations. Using this method
and high-resolution simulations, we make it clear whether tidal
detonation occurs during a WD~TDE. To come to the point, tidal
detonation occurs. Thus, we investigate its nucleosynthesis. It is the
first time that numerical simulations demonstrate tidal detonation of
a WD triggered by shock compression.

The paper is structured as follows. In section~\ref{sec:method}, we
describe our method which we validate in
Appendix~\ref{sec:validation}. In section~\ref{sec:results}, we show
our simulation results. We discuss about nucleosynthesis of this
WD~TDE in section~\ref{sec:discussion}. Finally, we summarize our
paper in section~\ref{sec:summary}. In this paper, we adopt CGS units
unless specified.

\section{Method}
\label{sec:method}

We follow tidal detonation of WD~TDEs as follows. We perform 3D SPH
simulations to follow overall WD~TDEs. In order to make initial
conditions of 1D mesh simulations, we extract WD structure along with
$z$-direction from the 3D SPH results. Finally, we perform 1D mesh
simulations. In subsection~\ref{sec:3d_sph_simulation}, we present our
method of 3D SPH simulations. In
subsection~\ref{sec:1d_initial_conditions}, we show how to make 1D
initial conditions. In subsection~\ref{sec:1d_mesh_simulation}, we
describe our method of 1D mesh simulations.

\subsection{3D SPH simulation}
\label{sec:3d_sph_simulation}

Our SPH method is similar to those in \cite{2017ApJ...839...81T}. We
adopt the vanilla-ice SPH equations for our SPH code. Our SPH kernel
function is Wendland $C^2$ Kernel
\citep{wendland1995piecewise,2012MNRAS.425.1068D}. A given particle
has neighbor particles of about $120$. Our artificial viscosity is the
same as proposed by \cite{1997JCoPh.136..298M}, and dependent on the
strength of a shock wave \citep{1997JCoPh.136...41M}. We suppress the
viscosity from shear motion, using Balsara's switch
\citep{1995JCoPh.121..357B}. We calculate gravitational forces among
particles with adaptive gravitational softening
\citep{2007MNRAS.374.1347P}. We parallelize our SPH code on
distributed-memory systems using FDPS \citep{2016PASJ...68...54I}, and
speed up explicit AVX instructions
\cite[e.g.][]{2012NewA...17...82T,2013NewA...19...74T}. We adopt the
Helmholtz equation of state (EoS) without Coulomb corrections
\citep{2000ApJS..126..501T}. We do not couple our SPH code with
nuclear reaction networks.

We choose an initial condition of a 3D SPH simulation as follows. Our
WD model has $0.45M_\odot$ and pure He composition, which is nearly
the upper limit of He~WD mass \citep{2017MNRAS.470.4473P}. It has no
spin. The numbers of SPH particles for the WD (hereafter $N_{\rm
  sph}$) are $4.7$, $9.4$, $19$, $38$, $75$, $150$, and $300$
millions, hereafter called $4.7$M, $9.4$M, $19$M, $38$M, $75$M,
$150$M, and $300$M, respectively. We relax the configurations of SPH
particles in the same way as \cite{2015ApJ...807...40T}. We
approximate IMBH gravitational potential as Newton potential. The IMBH
mass is $300M_\odot$. The orbit of the WD is parabolic around the
IMBH. The penetration factor $\beta$ is $7$. The IMBH does not irrupt
into nor swallow the WD. This is true even if we consider general
relativity for the IMBH gravity, using a generalized Newtonian
potential \citep{2013MNRAS.433.1930T}. An IMBH with $300M_\odot$
permits the closest encounter of a WD without irrupting into and
swallowing the WD among IMBHs \citep{2017arXiv170505526K}.  The
initial distance between the WD and IMBH is twice the tidal disruption
radius. We use $\tthr$ as the time from the starting time of a 3D
simulation. Then, the WD passes the pericenter during $\tthr=6$~s to
$\tthr=7$~s after 3D simulations start in all the $N_{\rm sph}$ cases.

\subsection{1D initial conditions}
\label{sec:1d_initial_conditions}

We make 1D initial conditions, extracting density and velocity
profiles in the $z$-direction from a WD whose evolution is followed by
a 3D SPH simulation. Hereafter, we call a portion of WD structure
extracted along the $z$-direction ``$z$-column'', and velocity in the
$z$-direction ``$z$-velocity''. We should minimize difference of
density and $z$-velocity evolution between 1D and 3D simulations. The
difference comes from 3D effects, such as a tidal field. In order to
minimize the 3D effects, we devise how to choose the time and place of
$z$-columns of the WD. If we follow 1D evolution for a long term, the
3D effects become significant. Thus, we should follow 1D evolution for
as short a term as possible. For this purpose, we start a 1D mesh
simulation just before tidal detonation is likely to occur. The tidal
detonation is triggered by a shock wave. The shock wave is formed from
steepening of a pressure wave generated by bounce of WD materials on
the $x$-$y$ plane. In order for the pressure wave to steepen into the
shock wave, relative velocity between WD materials on $x$-$y$ plane
and on the WD surface is supersonic \citep{2008A&A...481..259B}. Thus,
we extract $z$-columns in which WD materials on the surface approach
the $x$-$y$ plane at speed of Mach $4$ before the bounce. Moreover, we
select the densest $z$-columns among the above $z$-columns, since such
$z$-columns are easy to be detonated. In
Appendix~\ref{sec:compare1dWith3d}, we show our method minimizes 3D
effects.

\begin{figure}[ht!]
  \plotone{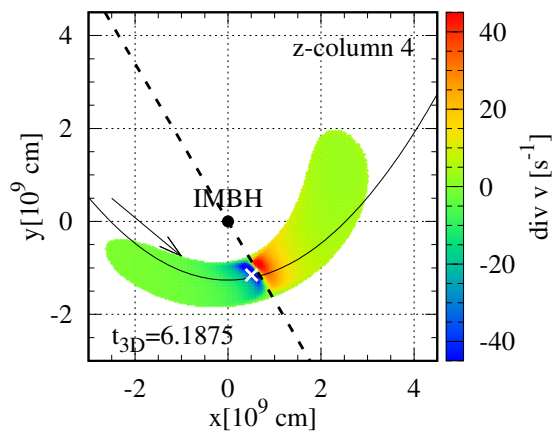}
  \caption{Divergence of velocity on the $x$-$y$ plane at the
    indicated time in 3D SPH simulation for the $N_{\rm sph}=38$M
    case. The IMBH is located at the coordinate origin. The solid
    curves show the orbit of the WD on the assumption that the WD is a
    point mass, and the arrows indicate the direction of the WD orbit.
    The dashed line in each panel ($y=-1.7x$) is a boundary dividing
    the WD into shrinking and expanding parts. White crosses indicate
    $z$-column~4, whose density is the highest among extracted
    $z$-columns. \label{fig:extract1dFrom3d_divv}}
\end{figure}

\begin{figure}[ht!]
  \plotone{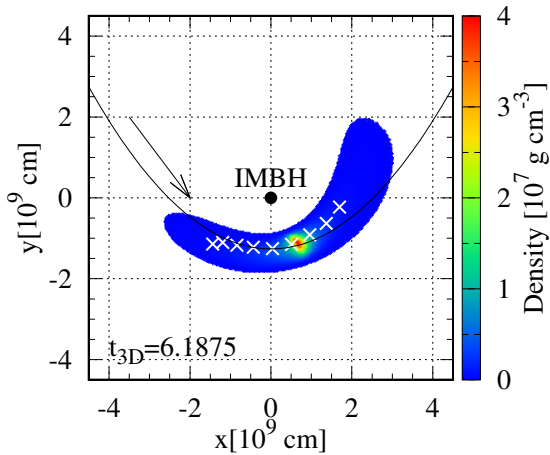}
  \caption{Density on the $x$-$y$ plane at $\tthr=6.1875$~s in 3D SPH
    simulation for the $N_{\rm sph}=38$M case. The IMBH is located at
    the coordinate origin. White crosses indicate $z$-columns~1 - 9 at
    $\tthr=6.1875$~s from right to
    left. \label{fig:extract1dFrom3d_dens}}
\end{figure}

We apply the above method to the $N_{\rm sph}=38$M case. For the other
$N_{\rm sph}$ cases, we extract $z$-columns whose time and place are
the same as those in the $N_{\rm sph}=38$M case, in order to perform a
convergence check with different $N_{\rm sph}$ of 3D SPH simulations
(see Appendix~\ref{sec:check3dResolution}). We extract $9$ $z$-columns
from 3D SPH simulation for each $N_{\rm sph}$ case. We refer to these
$z$-columns as $z$-columns~1, 2, $\dotsc$, and 9 in order from the
front in the orbital direction of the WD. As an example, we show
$z$-column~4 extracted from the $N_{\rm sph}=38$M case in
Figure~\ref{fig:extract1dFrom3d_divv}. The $z$-column has the highest
density among extracted $z$-columns. The $z$-column is located in a
shrinking region close to boundaries between shrinking and expanding
regions. Therefore, the $z$-column is located just before bounce.

Figure~\ref{fig:extract1dFrom3d_dens} shows the positions extracted
$z$-columns at $\tthr=6.1875$~s. From right to left, $z$-columns~1 - 9
are indicated by white crosses. Note that we do extract $z$-column~4
at this time, however we do not extract the other $z$-columns at this
time. We extract $z$-column~$x$ at $\tthr=6 + (x-1)/16$~s. These
$z$-columns are located in a shrinking region close to boundaries
between shrinking and expanding regions, similarly to
$z$-column~4. Since $z$-column~4 is closest to the maximum density
point of the WD among these $z$-columns, $z$-columns~1 - 3 precede the
maximum density point, and $z$-columns~5 - 9 follow the maximum
density point. Materials preceding $z$-column~1 have $0.05M_\odot$,
and materials following $z$-column~9 have $0.01M_\odot$. This means
that these $z$-columns cover about $90$~\% of the WD.

\begin{figure}[ht!]
  \plotone{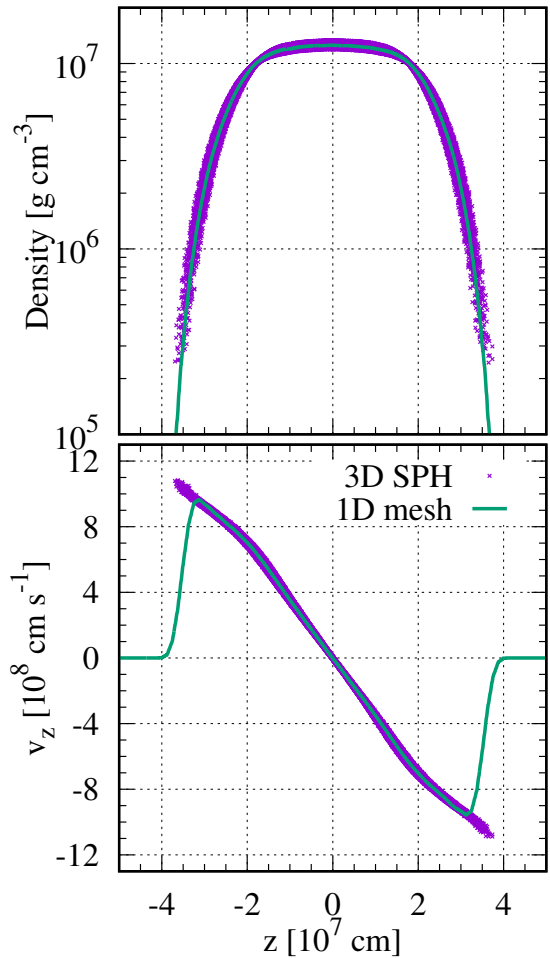}
  \caption{Profiles of density and $z$-velocity in $z$-column~4 in 3D
    SPH simulation for the $N_{\rm sph}=300$M case. Solid curves show
    1D profiles used for 1D initial conditions. Crosses indicate
    density and $z$-velocity of SPH particles in the 3D SPH
    simulation. \label{fig:compare1dWith3d_init}}
\end{figure}

In order to make 1D initial conditions, we calculate density and
$z$-velocity of the $z$-columns, using SPH kernel interpolation. We do
not use temperature of 3D SPH simulations, and set temperature to be
$10^6$~K. Although we set temperature to $10^5$~K, $10^6$~K, and
$10^7$~K, the temperature does not affect results of 1D mesh
simulations.

For the $N_{\rm sph}=300$M case, we indicate the calculated density
and $z$-velocity by solid curves in
Figure~\ref{fig:compare1dWith3d_init}. As reference, we also plot
density and $z$-velocity of SPH particles in $z$-column~4. The density
and $z$-velocity profiles for the 1D initial condition are in a good
agreement with those in the 3D SPH simulation, except the edge of the
$z$-column. In Appendix~\ref{sec:investigateEdge}, we describe the
discrepancy at the edge does not affect the emergence of tidal
detonation.

\subsection{1D mesh simulation}
\label{sec:1d_mesh_simulation}

We use the FLASH code \citep{2000ApJS..131..273F} for 1D mesh
simulations. The FLASH code is an Eulerian code. We use uniform mesh,
switching off adaptive mesh refinement. We adopt the piecewise
parabolic method \citep{1984JCoPh..54..174C} for the gas hydrodynamic
solver. Our EoS and nuclear reaction networks are the Helmholtz EoS
and Aprox13, respectively. Our timestep criterion is the minimum value
of the hydrodynamics timestep and nuclear reaction timestep. The
hydrodynamic timestep is $10$~\% of the Courant-Friedrichs-Lewy number
and the nuclear reaction timestep is $1$~\% of the ratio of the
specific internal energy to the specific nuclear energy-generation
rate. We do not consider self gravity of fluids. In most of 1D mesh
simulations, we do not include the IMBH gravity, however we partly
include the IMBH gravity in order to investigate its effect
(see Appendix~\ref{sec:investigateGravity}).

We set up calculation domain as follows. The domain geometry is
Cartesian. We have four cases of the domain range: $0 \le
z/10^8\mbox{cm} \le 0.50$, $|z/10^8\mbox{cm}| \le 0.5$,
$|z/10^8\mbox{cm}| \le 1.0$, and $|z/10^8\mbox{cm}| \le 2.0$. The
number of meshes is $6400$ in all the cases. Therefore, the mesh sizes
are $0.78 \times 10^4$~cm, $1.56 \times 10^4$~cm, $3.13 \times
10^4$~cm, and $6.25 \times 10^4$~cm. We adopt the domain range of
$|z/10^8\mbox{cm}| \le 4.0$ unless specified, and adopt the other
ranges for resolution study (see
Appendix~\ref{sec:check1dResolution}). The domain fully covers 1D WD
structure in the cases of the domain ranges of $|z/10^8\mbox{cm}| \le
1.0$, $|z/10^8\mbox{cm}| \le 2.0$, and $|z/10^8\mbox{cm}| \le 4.0$. In
these cases, the boundary condition is the outflow boundary at both
the edges of the domain. In the case of the domain range of $0 \le
z/10^8\mbox{cm} \le 0.50$, the domain covers only half the 1D WD
structure with $z \ge 0$. Thus, the boundary condition at $z=0$ is the
reflection boundary, and the boundary condition at the other edge is
the outflow boundary.

We use $\tone$ as the time from the starting time of a 1D mesh
simulation. The relation between $\tone$ and $\tthr$ is
$\tone=\tthr-(6+(x-1)/16)$~s for $z$-column~$x$.

We use initial conditions extracted from 3D SPH simulation with
$N_{\rm sph}=300$M unless specified.

\section{Results}
\label{sec:results}

\begin{figure*}[ht!]
  \plotone{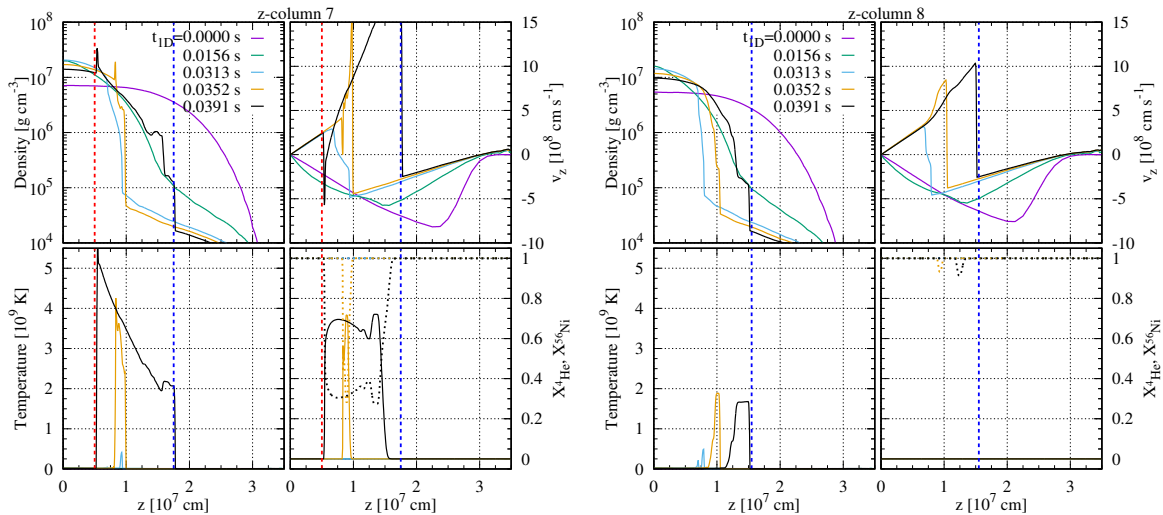}
  \caption{Time evolution of density, $z$-velocity, temperature, and
    nuclear composition profiles in $z$-column~7 (left four panels)
    and 8 (right four panels). The mass fractions of $^{4}$He and
    $^{56}$Ni are indicated by dotted and solid curves,
    respectively. Vertical dotted blue lines indicate forward shock
    waves, and vertical dotted red lines indicate a reverse shock wave
    accompanying a detonation wave. \label{fig:sample1dEvolution}}
\end{figure*}

We show success and failure cases of tidal detonation in order to
identify the emergence of tidal detonation easily. First, we present a
failure case. The right four panels of
Figure~\ref{fig:sample1dEvolution} show the time evolution of profiles
of density, $z$-velocity, temperature, and nuclear compositions of
$z$-column~8. At the beginning, the WD materials shrink in the
$z$-direction. At $\tone \sim 0.0234$~s, these materials bounce back,
and a pressure wave is generated. At $\tone \sim 0.0313$~s, the
pressure wave steepens into a shock wave. The shock wave is located at
$z \sim 1.5 \times 10^7$~cm at $\tone \sim 0.0391$~s, and indicated by
vertical dotted blue lines. The shock wave raises temperature of a
part of the materials. The temperature rise ignites nuclear
reactions. However, this nuclear reactions burn only small amounts of
$^{4}$He, and cease soon.

In this case, tidal detonation fails. This is because the shock wave
heats too a small region to trigger a detonation wave. The region
heated by the shock wave has $\sim 2 \times
10^6$~g~cm$^{-3}$. According to \cite{2013ApJ...771...14H}, detonation
in pure He composition arises only from a hotspot with size of
$>10^6$~cm if the density is $\sim 2 \times
10^6$~g~cm$^{-3}$. However, the size of the heated region is much
smaller than $10^6$~cm.

Next, we introduce a success case of tidal detonation. The left four
panels of Figure~\ref{fig:sample1dEvolution} show the time evolution
of $z$-column~7. The first-half evolution is similar to that of
$z$-column~8. At the initial time, the WD materials shrink. They
bounce back at $\tone \sim 0.0234$~s, and a pressure wave is
generated. At $\tone \sim 0.0313$~s, the pressure wave steepens into a
shock wave. The shock wave raises temperature of a part of the
materials. After that, the evolution is different from that of
$z$-column~8. The temperature rise triggers explosive nuclear
reactions. Since the nuclear reactions rapidly expand the materials,
they generate both forward and reverse shock waves. The forward and
reverse shock waves can be respectively seen at $z \sim 1.75 \times
10^7$~cm (vertical dotted blue lines) and $0.50 \times 10^7$~cm at
$\tone=0.0391$~s (vertical dotted red lines) in the $z$-velocity panel
of Figure~\ref{fig:sample1dEvolution} for $z$-column~7. The reverse
shock wave accompanies a detonation wave. In fact, behind the reverse
shock wave, large amounts of $^{4}$He and $^{56}$Ni are depleted and
yielded, respectively. Note that the reverse shock wave moves leftward
in Figure~\ref{fig:sample1dEvolution}.

In this case, tidal detonation succeeds. This is because the shock
wave heats a region large enough to trigger a detonation wave. The
region heated by the shock wave has $\sim 5 \times
10^6$~g~cm$^{-3}$. According to \cite{2013ApJ...771...14H}, detonation
in pure He composition arises only from a hotspot with size of
$>10^5$~cm if the density is $\sim 5 \times 10^6$~g~cm$^{-3}$. The
size of the heated region is much larger than $10^5$~cm.

The evidence of tidal detonation is the presence of a reverse shock
wave. A reverse shock wave can be easily seen in a $z$-velocity
profile.

\begin{figure*}[ht!]
  \plotone{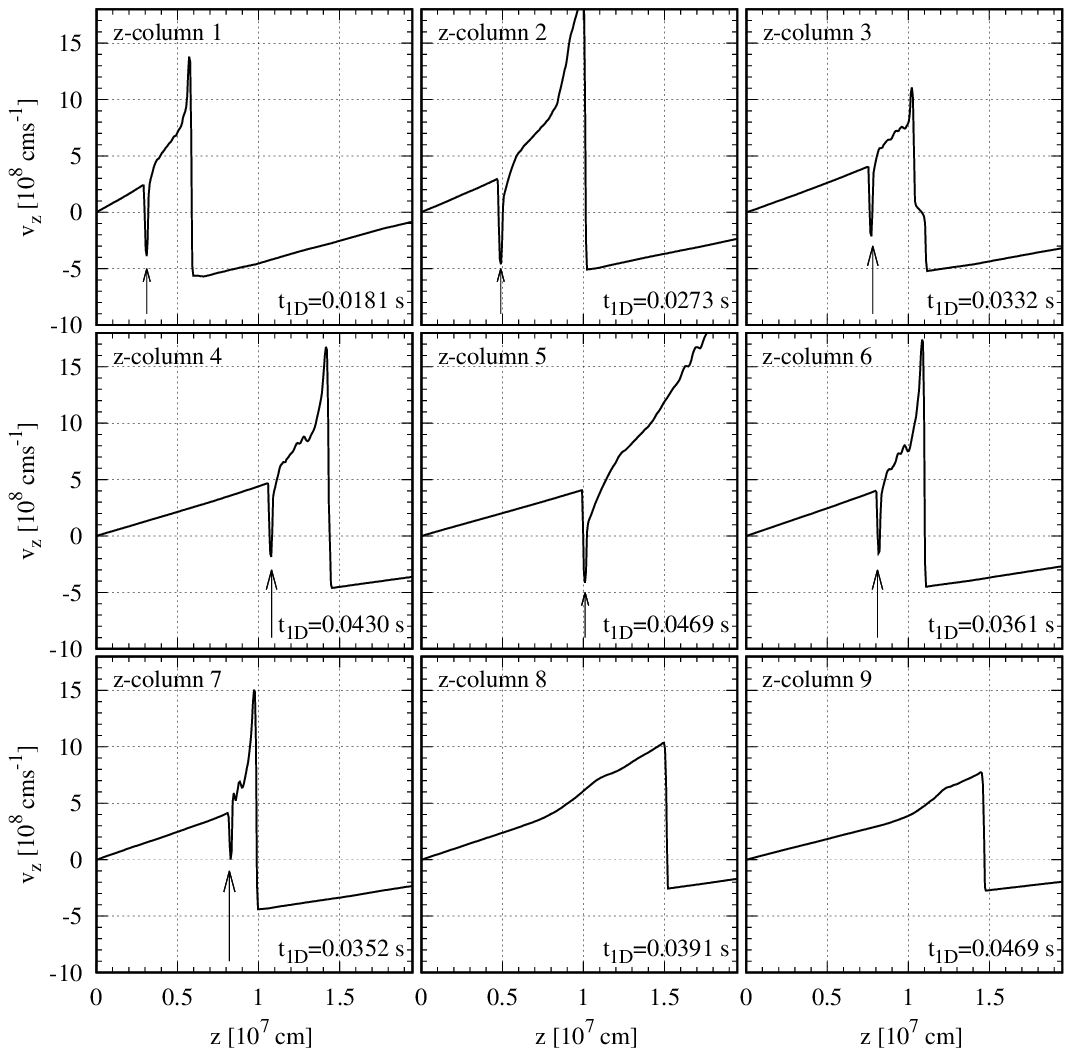}
  \caption{Profiles of $z$-velocity in $z$-columns~1 - 9 just after
    pressure waves steepen into shock waves. Arrows indicate reverse
    shock waves accompanying detonation
    waves. \label{fig:viewReverseShock_free}}
\end{figure*}

Figure~\ref{fig:viewReverseShock_free} shows the $z$-velocity profiles
for $z$-columns~1 - 9 just after pressure waves steepens into shock
waves. We can see reverse shock waves of $z$-columns~1 - 7 indicated
by arrows. In other words, tidal detonation occurs in $z$-columns~1 -
7, and does not in $z$-columns~8 - 9. The difference between the
former and latter $z$-columns is the distances from the IMBH. As seen
in Figure~\ref{fig:extract1dFrom3d_dens}, the leading part of the WD
tends to pass more closely to the IMBH than the trailing
part. Therefore, the former $z$-columns are closer to the IMBH (more
compressed) than the latter $z$-columns. We expect the materials
preceding $z$-column~1 succeed in tidal detonation, and the materials
following $z$-column~9 fail in tidal detonation.

\begin{figure*}[ht!]
  \plotone{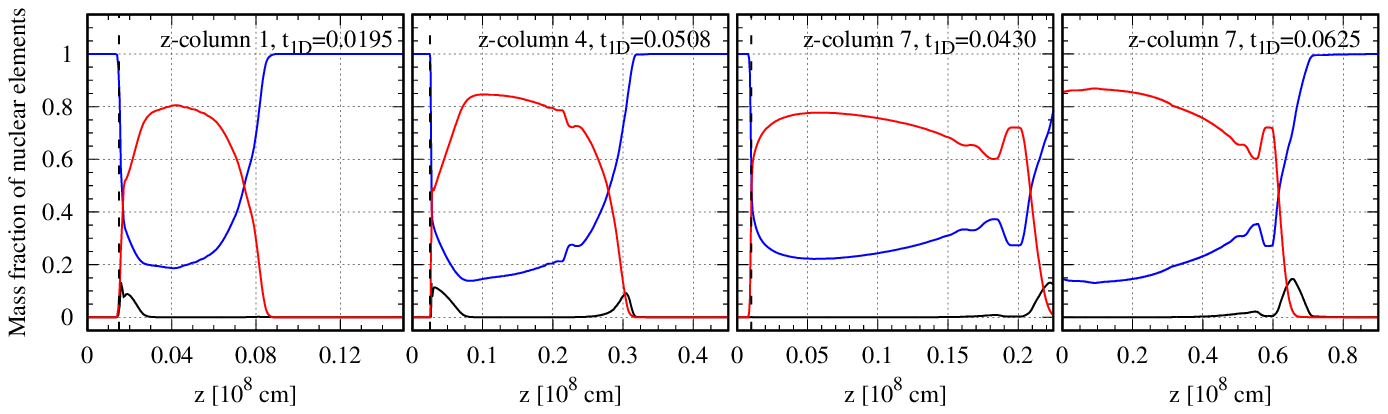}
  \caption{Profiles of nuclear elements in $z$-columns~1, 4 and
    7. Blue, black, and red curves indicate mass fractions of
    $^{4}$He, Si group elements ($^{28}$Si, $^{32}$S, $^{36}$Ar,
    $^{40}$Ca, and $^{44}$Ti), and $^{56}$Ni, respectively. Dashed
    lines indicate the positions of detonation waves. There is no
    dashed line in the right panel, since a detonation wave has
    reached the orbital plane ($z=0$), and nuclear reactions have been
    already finished.  \label{fig:viewNucleosynthesis}}
\end{figure*}

Figure~\ref{fig:viewNucleosynthesis} shows the profiles of nuclear
elements tidal detonation yields in $z$-columns~1, 4, and 7. The
nuclear reactions have been already finished in $z$-column~7 at
$\tone=0.625$~s. Most of materials have been burned by this time. It
seems that large amounts of unburned materials are left at $z > 0.7
\times 10^8$~cm. However, mass at $z > 0.7 \times 10^8$~cm is much
smaller than the total mass, since density at $z > 0.7 \times 10^8$~cm
is much smaller than at $z < 0.7 \times 10^8$~cm. The nuclear
reactions synthesize $80$~\% of $^{56}$Ni, and leave $20$~\% of
$^{4}$He in mass. There are small amounts of Si group elements
($^{28}$Si, $^{32}$S, $^{36}$Ar, $^{40}$Ca, and $^{44}$Ti), and their
mass fraction is $0.3$~\%. The reason for the small amounts of Si
group elements is that detonated materials have high density ($\gtrsim
10^7$~g~cm$^{-3}$). Note that Si group elements are synthesized when a
detonation wave proceeds in a region with density of $\lesssim
10^6$~g~cm$^{-3}$.

In the cases of $z$-columns~1 and 4, we have not finished nuclear
reactions. These 1D mesh simulations are largely
time-consuming. Timestep becomes too small just before the detonation
wave reaches the orbital plane ($z=0$) due to nuclear reactions. Here,
we make a conjecture about nucleosynthesis in $z$-columns~1 and 4 from
the results of $z$-column~7. In the second right panel of
Figure~\ref{fig:viewNucleosynthesis}, we show nuclear components in
$z$-column~7 just before a detonation wave reaches the orbital
plane. The nuclear components in $z$-columns~1, 4, and 7 are similar
just before the detonation waves reach the orbital plane. Therefore,
we expect nuclear components in these $z$-columns will be similar when
nuclear reactions have finished.

The 1D mesh simulation in the case of $z$-column~7 has been finished
in the following reason. Since $z$-column~7 follows $z$-columns~1 and
4, the former is less compressed than the latter. The former density
at $z=0$ is smaller. The nuclear reactions in $z$-column~7 are less
active than in $z$-columns~1 and 4. Therefore, timestep in the case of
$z$-column~7 does not become as small as in the case of $z$-columns~1
and 4.

\section{Discussion}
\label{sec:discussion}

We estimate overall nucleosynthesis in our WD~TDE model. We also
examine 3D effects for the nucleosynthesis.

\begin{figure}[ht!]
  \plotone{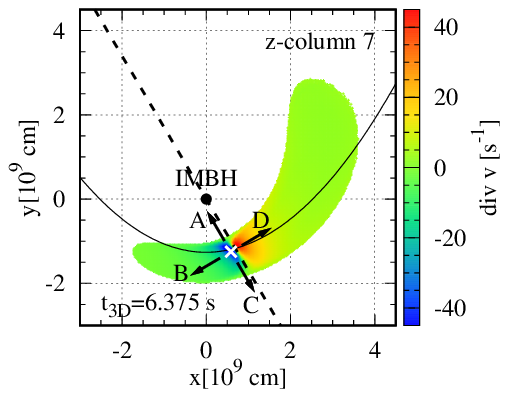}
  \caption{Divergence of velocity on the $x$-$y$ plane at the
    indicated time in 3D SPH simulation for the $N_{\rm sph}=38$~M
    case. The white cross indicates $z$-column~7. Four arrows from
    $z$-column~7 are depicted. Arrow~A points to the direction of the
    IMBH. Arrow~D points to the traveling direction of the
    WD. Arrows~C and B point to the inverse directions of arrows~A and
    D, respectively.
\label{fig:discussTraverseDet}}
\end{figure}

We consider whether a detonation wave proceeds in the direction of the
$x$-$y$ plane. Figure~\ref{fig:discussTraverseDet} shows the position
of $z$-column~7 (the white cross) and four arrows from
$z$-column~7. These arrows points to four directions on the $x$-$y$
plane. Hereafter, we refer to the direction of arrow~X as the
X-direction. The detonation wave from $z$-column~7 would not proceed
in the A-, C-, and D-directions. The $z$-columns in the D-direction
have been already detonated. The $z$-columns in the A- and
C-directions are detonated simultaneously with $z$-column~7.

The detonation wave from $z$-column~7 would proceed in the
B-direction, if no detonation wave is generated from $z$-columns in
the B-direction. If detonation waves are generated from $z$-columns in
the B-direction, the detonation wave from $z$-column~7 would not
proceed in the B-direction. Detonation waves from $z$-columns in the
B-direction would be spontaneously generated before the detonation
wave from $z$-column~7 reaches these $z$-columns. This is because the
speed of the detonation wave ($\sim 10^9$~cm~s$^{-1}$) is much slower
than the orbital velocity of the WD ($\sim 8 \times
10^9$~cm~s$^{-1}$).

Based on the above consideration, we can divide $z$-columns of a WD
into two types. The first type is $z$-columns in which tidal
detonation arises, such as $z$-columns~1 - 7. The second type is
$z$-columns which tidal detonation arising in other $z$-columns
reaches. Since tidal detonation succeeds in $z$-columns~7, and fails
in $z$-columns~8, we assume that $z$-columns preceding $z$-column~7
are the first type, and that $z$-columns following $z$-column~7 are
the second type. Note that leading parts are easier to be detonated
than trailing parts, as described in section~\ref{sec:results}.

We discuss about the first type of $z$-columns. All these $z$-columns
would be detonated independently of each other. These $z$-columns have
mass of $0.37M_\odot$. Their nuclear compositions would be similar to
those of $z$-columns~1, 4 and 7 (see
Figure~\ref{fig:viewNucleosynthesis}). Then, their nuclear
compositions would be $^{56}$Ni of $0.30M_\odot$ and $^{4}$He of
$0.07M_\odot$. There would be small amounts of Si group elements,
$\sim 0.001M_\odot$.

We examine the second type of $z$-columns. They would be detonated by
the detonation wave generated in $z$-column~7. From our simulation
results, we find the detonation wave traverses these $z$-columns when
these $z$-columns have density of $\sim
10^{5}$~g~cm$^{-3}$. Therefore, the detonation wave would yield large
amounts of Si group elements in these $z$-columns
\citep{2013ApJ...771...14H}. Since these $z$-columns have
$0.08M_\odot$, Si group elements of $0.08M_\odot$ would be
synthesized.

In summary, our WD~TDE model would yield $^{56}$Ni of $0.30M_\odot$
and Si group elements of $0.08 M_\odot$, and leave $^{4}$He of
$0.07M_\odot$. Since parts of them would be swallowed by the IMBH, all
of them would not contribute to the luminosity of our WD~TDE
model. However, we do not investigate the subsequent evolution of the
WD~TDE. We cannot follow the orbit of the WD debris accurately, since
we simplify the IMBH gravity as Newton gravity.

\section{Summary}
\label{sec:summary}

We assess whether tidal detonation occurs true in WD~TDEs. We choose a
He~WD with $0.45M_\odot$. We need prohibitively large calculation cost
to follow tidal detonation of a WD~TDE by 3D SPH simulations. Thus, we
combine 1D mesh simulations with 3D SPH simulations. For 1D mesh
simulations, we develop a method to extract 1D initial conditions from
3D SPH simulation data. Owing to the method, we can follow 1D
evolution, not annoyed by 3D effects, such as a tidal field.

We show tidal detonation arises by shock heating. We emphasize it is
the first time that numerical simulations demonstrate tidal detonation
of a WD triggered by shock heating. For this purpose, we perform
severe convergence checks with different $N_{\rm sph}$ of 3D SPH
simulations and with different space resolution of 1D mesh
simulations. Tidal detonation succeeds in $z$-columns preceding
$z$-column~7, and fails in $z$-columns following $z$-column~7. Leading
parts are easier to be detonated in the following reason. Leading
parts are more compressed by the IMBH than trailing parts, and tend to
have higher density. A detonation wave is easier to occur in a higher
density region, since nuclear reactions proceed more rapidly.

For $z$-columns preceding $z$-column~7, the detonation waves would
yield large amounts of $^{56}$Ni, since these $z$-columns have high
density, $\sim 10^7$~g~cm$^{-3}$. Materials in $z$-columns following
$z$-column~7 would be detonated by the detonation wave arising from
$z$-column~7. In these $z$-columns, large amounts of Si group elements
would be synthesized, since these $z$-columns have density $\sim
10^5$~g~cm$^{-3}$ when these $z$-columns receive the detonation
wave. Eventually, our WD~TDE model would synthesize $^{56}$Ni of
$0.30M_\odot$ and Si group elements of $0.08M_\odot$, and would leave
$^{4}$He of $0.07M_\odot$. Therefore, the WD~TDE could be observed as
a luminous thermonuclear transient comparable to type Ia supernovae.

\acknowledgments

Numerical computations were carried out on Cray XC30 at Center for
Computational Astrophysics, National Astronomical Observatory of
Japan, on Cray XC40 at Yukawa Institute for Theoretical Physics, Kyoto
University, and on Oakforest-PACS at Joint Center for Advanced High
Performance Computing. The software used in this work was in part
developed by the DOE NNSA-ASC OASCR Flash Center at the University of
Chicago. This research has been supported in part by MEXT program for
the Development and Improvement for the Next Generation Ultra
High-Speed Computer System under its Subsidies for Operating the
Specific Advanced Large Research Facilities, and by Grants-in-Aid for
Scientific Research (16K17656, 17H06360) from the Japan Society for
the Promotion of Science.

\software{FLASH \citep{2000ApJS..131..273F,2010ascl.soft10082F}}

\appendix

\section{Validation of our method}
\label{sec:validation}

\subsection{Comparison between 3D and 1D simulations}
\label{sec:compare1dWith3d}

\begin{figure*}[ht!]
  \plotone{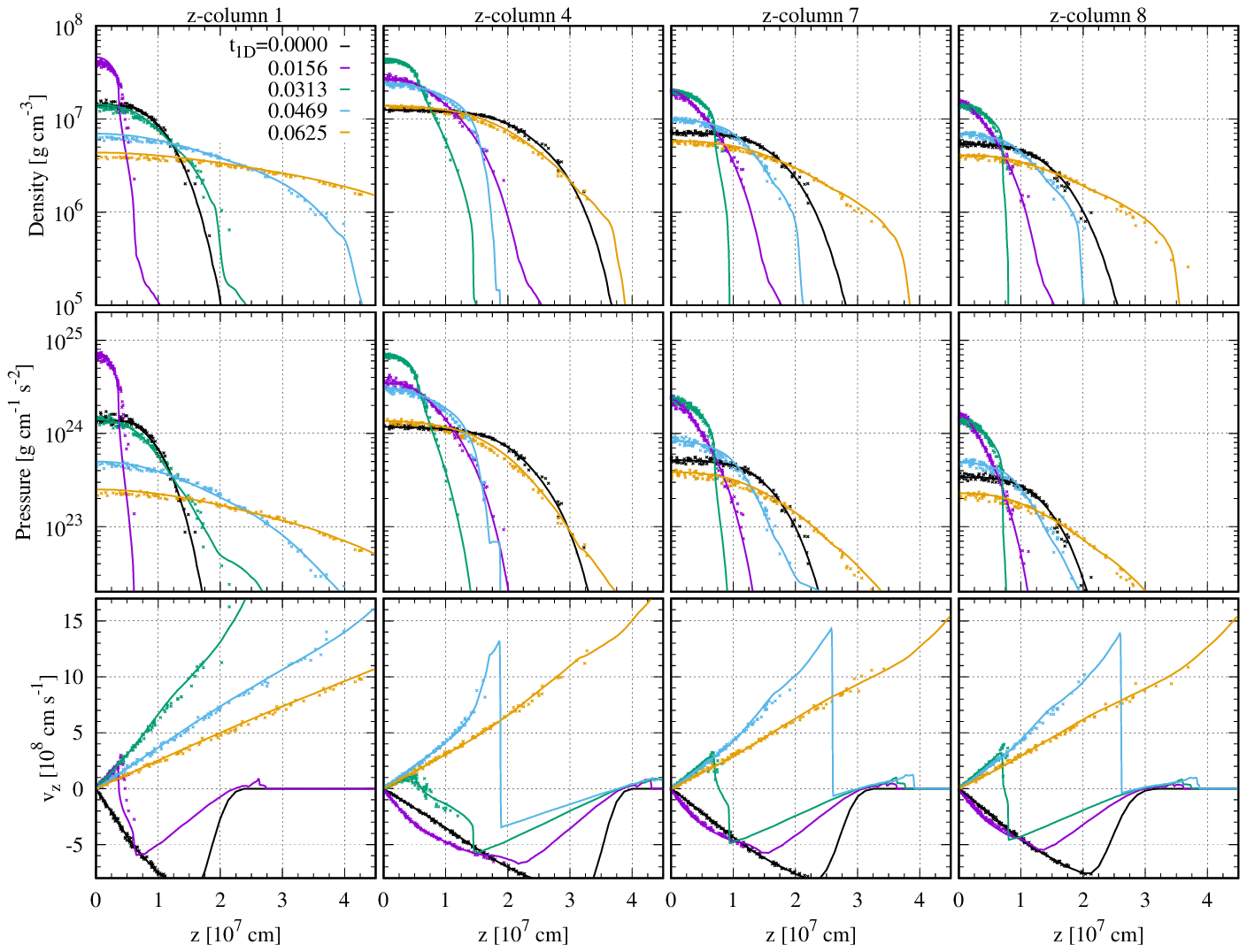}
  \caption{Density, pressure and $z$-velocity profiles in
    $z$-columns~1, 4, 7, and 8. Points indicate SPH particles sampled
    randomly from 3D SPH simulation for saving storage size. The
    number of the sampled particles for each $z$-column is about 300th
    of the number of all the particles in each $z$-column. Solid
    curves show the results of the 1D mesh simulation. For this
    comparison, we turn off nuclear reaction networks in the 1D mesh
    simulation. \label{fig:compare1dWith3d_evol}}
\end{figure*}

In Figure~\ref{fig:compare1dWith3d_evol}, we compare the time
evolution of $z$-columns~1, 4, 7, and 8 in the 3D SPH simulation of
$N_{\rm sph}=300$M with the time evolution of these $z$-columns in the
1D mesh simulation. For this comparison, we turn off nuclear reaction
networks in the 1D mesh simulation. We find a good agreement between
the 3D SPH and 1D mesh simulation results. Therefore, 3D effects, such
as a tidal field, are not significant during this time.

\subsection{Comparison between simulations with and without IMBH gravity}
\label{sec:investigateGravity}

\begin{figure*}[ht!]
  \plotone{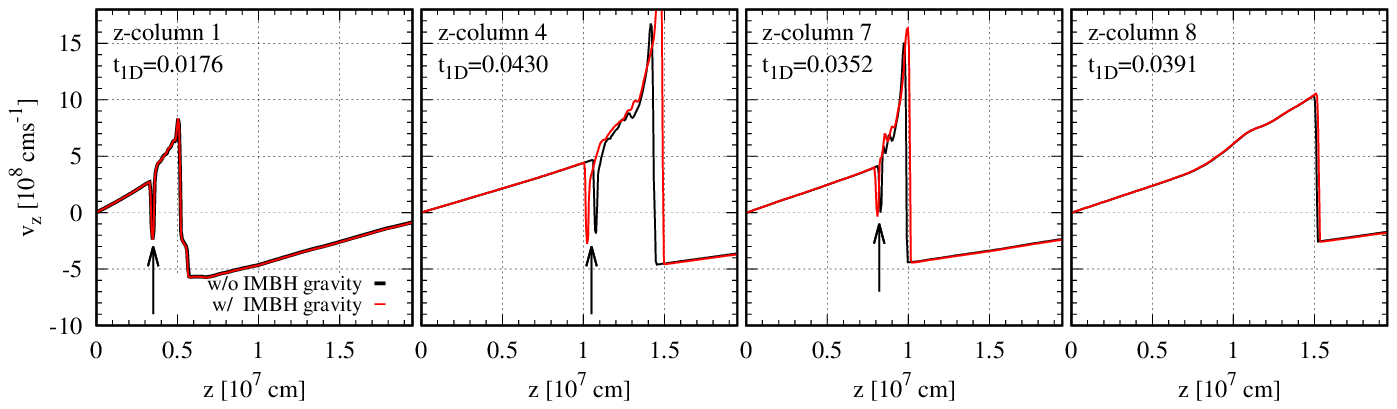}
  \caption{Profiles of $z$-velocity in $z$-columns~1, 4, 7, and 8 just
    after pressure waves steepen into shock waves. Red and black
    curves indicate with and without including the IMBH gravity in the
    $z$-direction, respectively. Arrows indicate reverse shock waves
    accompanying detonation waves. \label{fig:viewReverseShock_grav}}
\end{figure*}

We perform 1D mesh simulations, including the IMBH gravity in the
$z$-direction. We set separation between the $z$-columns and IMBH to
$1.45 \times 10^9$~cm (see Figures~\ref{fig:extract1dFrom3d_divv} and
\ref{fig:extract1dFrom3d_dens}). Figure~\ref{fig:viewReverseShock_grav}
shows the $z$-velocity profiles just after pressure waves steepen into
shock waves in the cases including the IMBH gravity. Tidal detonation
occurs in $z$-columns~1 - 7, and does not in $z$-columns~8 - 9, which
is the same as in the cases where we ignore the IMBH gravity. This is
because pressure gradients in the $z$-columns are much larger than the
IMBH gravity. For example, tidal detonation starts at $z \sim 10^7$~cm
and $\tone \sim 0.0352$~s in $z$-column~7 (see
Figure~\ref{fig:sample1dEvolution}). In that place, the pressure
gradient and IMBH gravity are, respectively, $(\partial P / \partial
z) / \rho \gtrsim 10^{10}$~cm~s$^{-2}$ (see the curves of
$\tone=0.0313$~s and $0.0469$~s of $z$-column~7 panels in
Figure~\ref{fig:compare1dWith3d_evol}) and $GM_{\rm IMBH}z/R^3 \sim
10^8$~cm~s$^{-2}$.

\subsection{Modeling of the WD edge}
\label{sec:investigateEdge}

As seen in the top panel of Figure~\ref{fig:compare1dWith3d_init},
density structure in the 1D initial condition is extrapolated below
$2 \times 10^5$~g~cm$^{-3}$. This extrapolation does not affect the
emergence of tidal detonation, since the tidal detonation emerges at a
region reflecting information of 3D SPH simulation. In $z$-column~4,
tidal detonation occurs where density is $> 10^6$~g~cm$^{-3}$. On the
other hand, materials in this $z$-column are compressed by a factor of
at most $5$ during its evolution. Therefore, the tidal detonation
emerges at materials whose density is initially $> 2 \times
10^5$~g~cm$^{-3}$. In other words, the tidal detonation occurs at a
region reflecting 3D SPH simulation results.

\begin{figure*}[ht!]
  \plotone{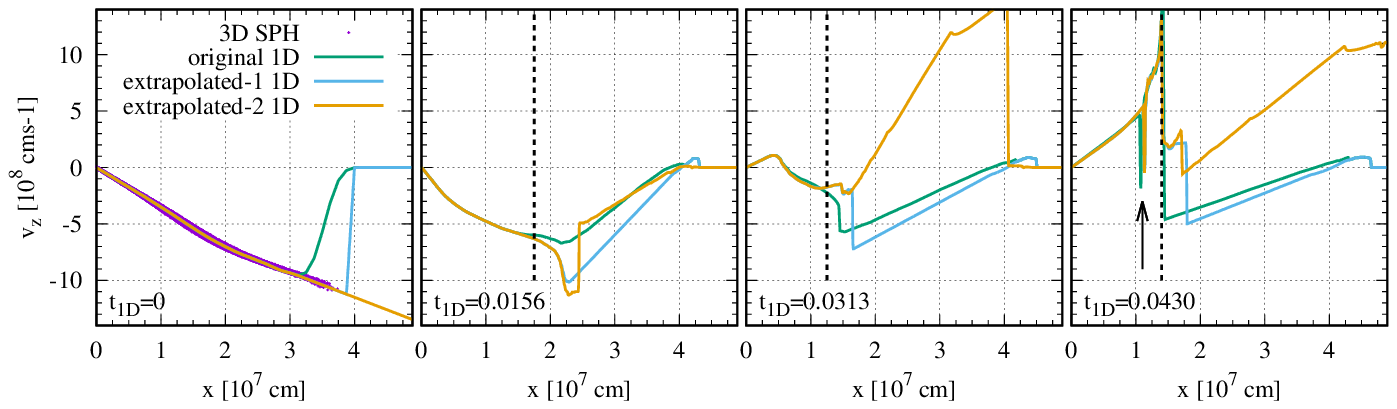}
  \caption{Time evolution of $z$-velocity profiles in $z$-column~4 for
    original and extrapolated 1D initial conditions. As a reference,
    we show the $z$-velocity profile in the 3D SPH simulation at
    $\tone=0$. The arrow indicates reverse shock waves accompanying
    detonation waves. \label{fig:checkVzSetup}}
\end{figure*}

As seen in the bottom panel of Figure~\ref{fig:compare1dWith3d_init},
$z$-velocity structure in the 1D initial condition is slightly
oversmoothed at the edge of the $z$-column. We investigate this
oversmoothing on the emergence of tidal detonation. Using $z$-velocity
structure in the 3D SPH simulation, we extrapolate $z$-velocity
structure for $z$-column~4 in two different ways. One is that we set
$z$-velocity to zero at $|z|=4 \times 10^7$~cm discontinuously, and
the other is that we increase the absolute value of $z$-velocity up to
the boundary of the 1D mesh simulation (see the left panel of
Figure~\ref{fig:checkVzSetup}). As seen in the second and third left
panels of Figure~\ref{fig:checkVzSetup}, the modeling of the WD edge
affects the edge evolution of $z$-velocity (on the right side of the
vertical dotted lines), however does not affects the internal
evolution of $z$-velocity which raises pressure waves steepening into
shock waves (on the left side of the vertical dotted lines). Finally,
we can see in Figure~\ref{fig:checkVzSetup} that tidal detonation
emerges in both cases of the extrapolated 1D initial conditions,
similarly to the original 1D initial conditions. Therefore, the
oversmoothing $z$-velocity at the edge of $z$-columns does not affect
the emergence of tidal detonation.

\subsection{1D resolution check}
\label{sec:check1dResolution}

\begin{figure*}[ht!]
  \plotone{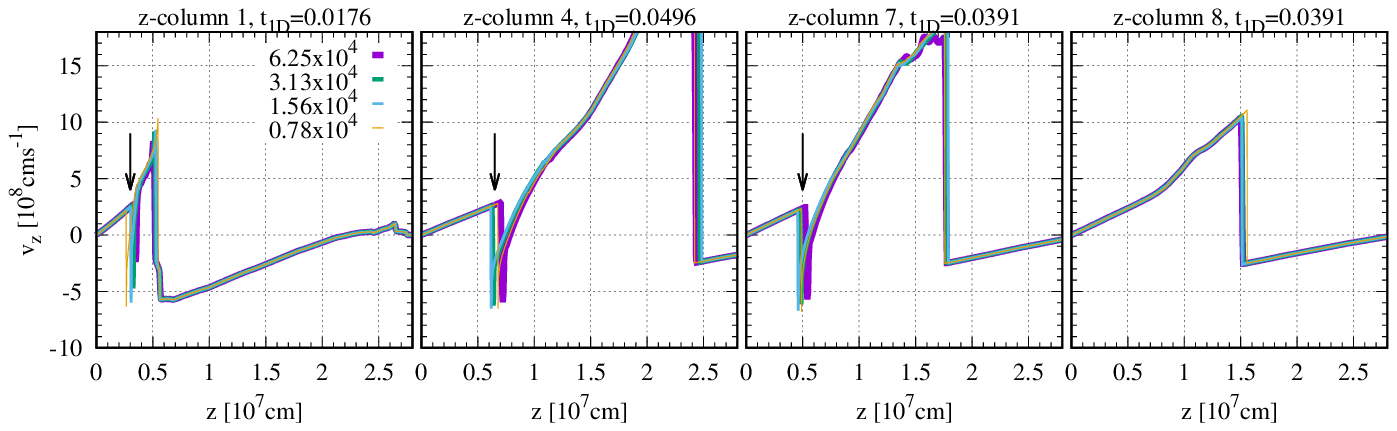}
  \caption{Profiles of $z$-velocity in $z$-columns~1, 4, 7, and 8 with
    different space resolution from $6.25 \times 10^4$~cm to $0.78
    \times 10^4$~cm in 1D mesh simulations. The time is just after
    tidal detonation (or just a shock wave) appears. Arrows indicate
    reverse shock waves accompanying detonation
    waves. \label{fig:check1dResolution}}
\end{figure*}

We perform a convergence check of 1D mesh simulations with different
space resolution for $z$-columns~1, 4, 7, and 8. We change the mesh
size from $6.25 \times 10^4$~cm to $0.78 \times 10^4$~cm. Note that
the mesh size sufficiently resolves the hotspot size ($\sim
10^5$~cm). Figure~\ref{fig:check1dResolution} shows the time evolution
of $z$-velocity profiles with different 1D space resolution. We find a
good agreement among the results of different 1D space resolution. We
conclude the results about the emergence of tidal detonation are
converged among different 1D space resolution.

\subsection{3D resolution check}
\label{sec:check3dResolution}

\begin{figure*}[ht!]
  \plotone{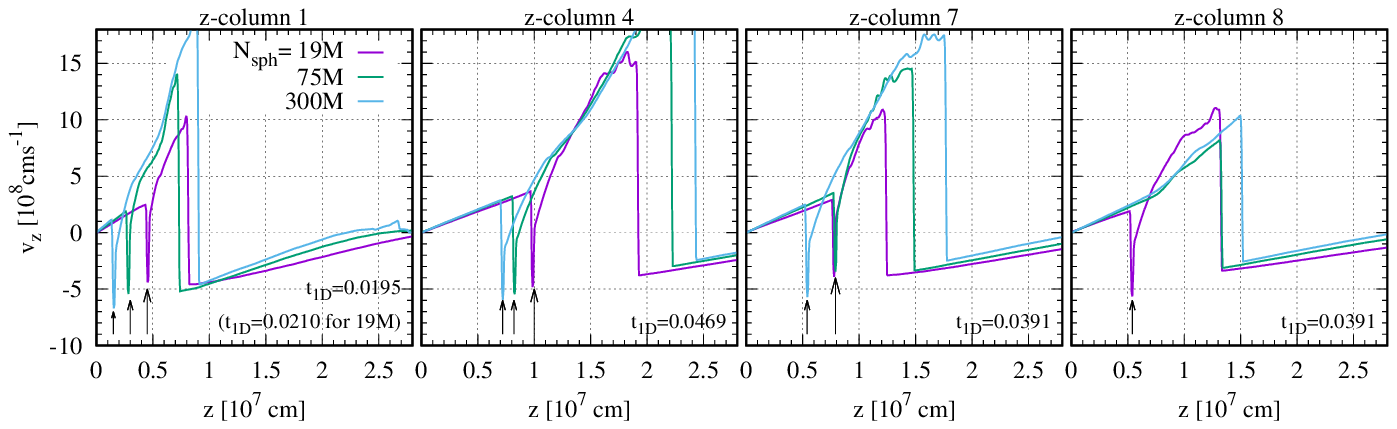}
  \caption{Profiles of $z$-velocity in $z$-columns~1, 4, 7, and 8 for
    3D SPH simulations with the $N_{\rm sph}=19$M, $75$M, and $300$M
    cases. Arrows indicate reverse shock waves accompanying detonation
    waves. \label{fig:view3dResolution}}
\end{figure*}

We perform a convergence check of 1D mesh simulations with different
$N_{\rm sph}$ in 3D SPH simulations. Note that this is {\it not} a
convergence check of space resolution in 1D mesh
simulations. Figure~\ref{fig:view3dResolution} shows the time
evolution of $z$-velocity profiles in $z$-columns~1, 4, 7, and 8 for
3D SPH simulations with $N_{\rm sph}=19$M, $75$M, and $300$M. We can
see the presence of reverse shock waves in all $N_{\rm sph}$ cases for
$z$-columns~1, 4 and 7. On the other hand, for $z$-column~8, a reverse
shock wave is present in the $N_{\rm sph}=19$M case, and absent in the
the $N_{\rm sph}=75$M and $300$M cases. Eventually, we find tidal
detonation succeeds in the $N_{\rm sph}<75$M cases, and fails in the
$N_{\rm sph} \ge 75$M cases for $z$-column~8.

\begin{figure*}[ht!]
  \plotone{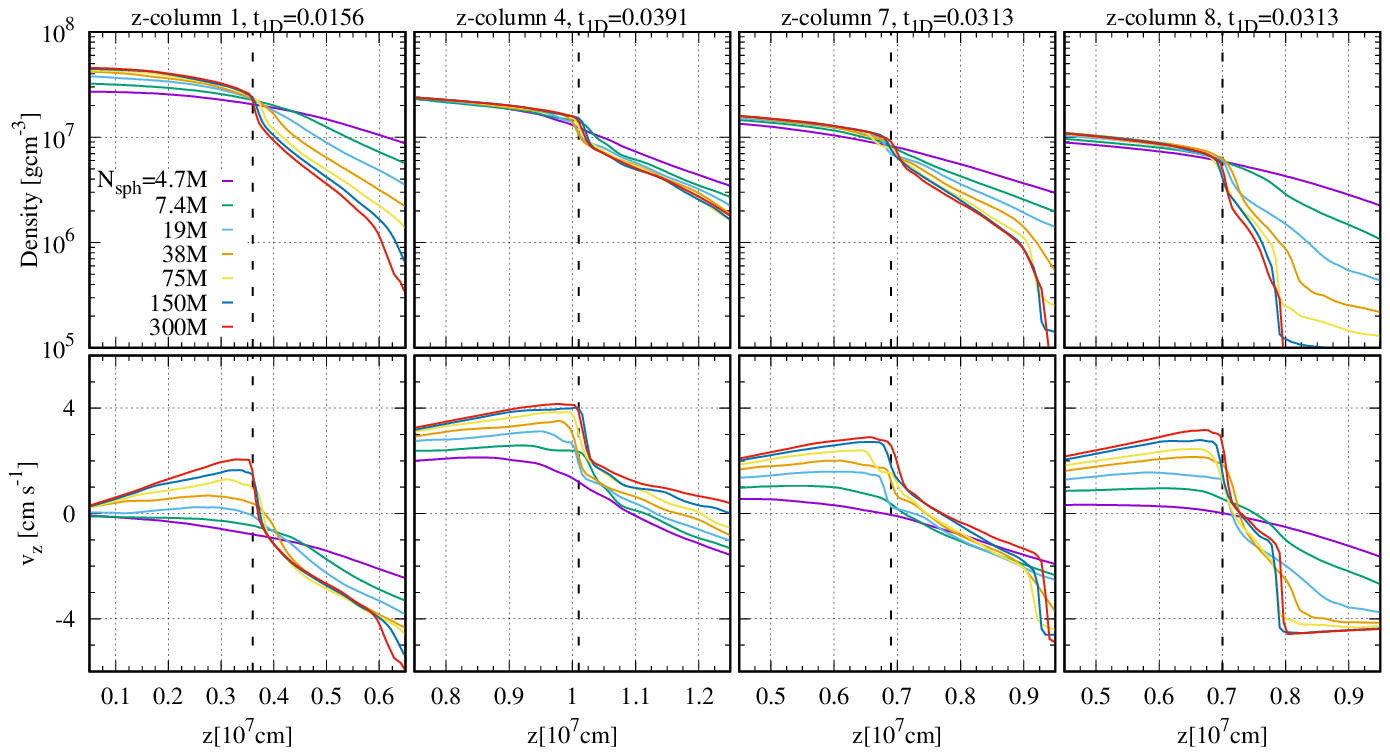}
  \caption{Density and $z$-velocity profiles in $z$-columns~1, 4, 7,
    and 8 just before shock waves appears. The 3D SPH resolution is
    $N_{\rm sph}=4.7$M, $9.4$M, $19$M, $38$M, $75$M, $150$M, and
    $300$M. Dashed lines indicate the position of a pressure wave to
    be a shock wave immediately.
    \label{fig:check3dResolution}}
\end{figure*}

In order to investigate the reason for the failure of tidal detonation
in large-$N_{\rm sph}$ cases, we show density and $z$-velocity
profiles in these $z$-columns just before shock waves appear in
Figure~\ref{fig:check3dResolution}. Just after this time, a pressure
wave indicated by dashed lines will steepen into a shock wave
immediately. Thus, the shock wave will heat the right-side region of
the dashed line after this time. In all the $z$-columns, we can see
density of the right-side region becomes smaller with $N_{\rm sph}$
increasing if $N_{\rm sph} < 75$M. Since nuclear reactions proceed
more rapidly with density higher, tidal detonation occurs more easily
in smaller $N_{\rm sph}$ cases. In all the $z$-columns, density
profiles in the $N_{\rm sph} \ge 75$M cases are the same. Therefore,
the 1D mesh simulations are converged in the range from $N_{\rm
  sph}=75$M to $N_{\rm sph}=300$M. In other words, tidal detonation
fails in $z$-column~8, and tidal detonation will succeed in
$z$-columns~1, 4 and 7 even if $N_{\rm sph}$ becomes infinite.

\begin{figure}[ht!]
  \begin{center}
    \includegraphics[width=7cm]{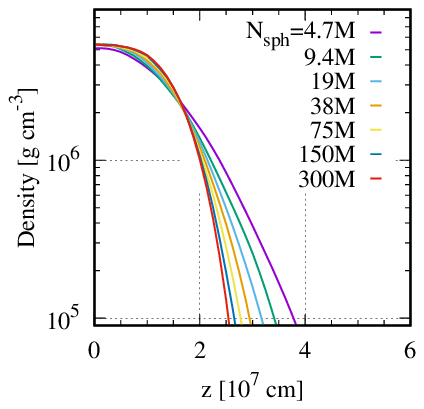}
  \end{center}
  \caption{Density profile in $z$-column~8 at the initial time, based
    on 3D SPH simulation with different $N_{\rm sph}$.
    \label{fig:check1dInitCondition}}
\end{figure}

The reason for the unconvergence in the $N_{\rm sph}<75$M cases is
that initial conditions of 1D mesh simulations in the $N_{\rm
  sph}<75$M cases are not converged, since the results of 3D SPH
simulations are not converged especially at the WD
edge. Figure~\ref{fig:check1dInitCondition} shows the initial
conditions of $z$-column~8 in 3D SPH simulations with $N_{\rm
  sph}=4.7$M to $300$M. At the edge of the $z$-column, density
decreases with $N_{\rm sph}$ increasing. Generally, SPH methods can
capture edge structure sharply with $N_{\rm sph}$ increasing, since
SPH kernel length becomes smaller with $N_{\rm sph}$
increasing. Therefore, density at the edge is overestimated, and tidal
detonation occurs falsely for smaller-$N_{\rm sph}$ cases.

\end{document}